# Electronic Localization in CaVO$_3$ Films via Bandwidth Control

Daniel E. McNally[1], Xingye Lu[1], Jonathan Pelliciari[1,†], Sophie Beck[2], Marcus Dantz[1], Muntaser Naamneh[1], Tian Shang[1,3], Marisa Medarde[4], Christof W. Schneider[5], Vladimir N. Strocov[1], Ekaterina V. Pomjakushina[4], Claude Ederer[2], Milan Radovic[1#], and Thorsten Schmitt[1]*

[1] Photon Science Division, Paul Scherrer Institut, CH-5232 Villigen PSI, Switzerland.

[2] Materials Theory, ETH Zurich, Wolfgang-Pauli Strasse 27, CH-8093 Zurich, Switzerland

[3] Institute of Condensed Matter Physics, Ecole Polytechnique Federale de Lausanne (EPFL), CH-1015 Lausanne, Switzerland

[4] Laboratory for Scientific Developments and Novel Materials, Paul Scherrer Institut, CH-5232 Villigen PSI, Switzerland

[5] Laboratory for Multiscale Materials Experiments, Paul Scherrer Institut, CH-5232 Villigen PSI, Switzerland

[†] Current address: Department of Physics, Massachusetts Institute of Technology, Cambridge, Massachusetts 02139, USA

* thorsten.schmitt@psi.ch
# milan.radovic@psi.ch

**Abstract:**

**Understanding and controlling the electronic structure of thin layers of quantum materials is a crucial first step towards designing heterostructures where new phases and phenomena, including the metal-insulator transition (MIT), emerge. Here, we demonstrate control of the MIT via tuning electronic bandwidth and local site environment through selection of the number of atomic layers deposited. We take $CaVO_3$, a correlated metal in its bulk form that has only a single electron in its $V^{4+}$ 3d manifold, as a representative example. We find that thick films and ultrathin films (≤ 6 unit cells, u.c.) are metallic and insulating, respectively, while a 10 u.c. $CaVO_3$ film exhibits a clear thermal MIT. Our combined X-ray absorption spectroscopy and resonant inelastic x-ray scattering (RIXS) study reveals that the thickness-induced MIT is triggered by electronic bandwidth reduction and local moment formation from $V^{3+}$ ions, that are both a consequence of the thickness confinement. The thermal MIT in our 10 u.c. $CaVO_3$ film exhibits similar changes in the RIXS response to that of the thickness-induced MIT in terms of reduction of bandwidth and V 3d – O 2p hybridization.**

**Introduction**

Deposition methods with unit cell thickness precision enable new approaches to control the degree of electronic localization in quantum materials. For instance, epitaxial strain changes the lattice parameters altering the potential landscape of the electrons, dimensionality tunes the electronic bandwidth and interfacing provides a charge reservoir. Thickness-dependent metal-insulator transitions (MIT) in thin films have been widely reported in recent years in a broad swathe of correlated 3d and 5d transition metal oxides [1-7]. Understanding and controlling the electronic structure of these quantum materials is a critical first step toward designing heterostructures where new emergent phases and phenomena, including MIT, have been widely predicted [8, 9]. The focus of our interest is to resolve the nature of the thickness-induced MIT, therefore, in this work we consider $CaVO_3$ as a model system.

Bulk $CaVO_3$ with a formal valence shell configuration of V3$d^1$ ($V^{4+}$ oxidation state) is a paramagnetic metal with enhanced Sommerfeld-Wilson and Kadowaki-Woods ratios suggesting electronic correlations are important [10]. However, as is often the case for metals, neither moderate doping nor pressure induce insulating or magnetic ground states [11,12]. In contrast to the bulk, enhanced electron correlation strength and a MIT have recently been reported for $CaVO_3$ films of 20 unit cells (u.c.) or less [6,13]. While these discoveries are promising for technological applications such as transparent conductors [13] and MottFETs [14,15], a complete understanding of the MIT in $CaVO_3$ has, until now, remained elusive.

Figure 1 illustrates that electronic bandwidth can control the degree of electron localization in 3$d^1$ perovskites $RXO_3$. In the example presented in the top panels, applicable for bulk-like materials, when the orthorhombic distortion is increased the bandwidth decreases by up to 40 % and a MIT occurs [16]. However, realising this MIT requires the substitution of both R and X ions making it very difficult to continuously control in practice. In our work presented here we show that the bandwidth can be controlled continuously by up to 40 % in $CaVO_3$ simply by choosing the number of atomic layers deposited on a $SrTiO_3$ substrate as summarized in the bottom panels of Figure 1 and outlined in detail in the rest of this work.

## Results

**Metal-insulator transition in thin film CaVO$_3$.**

Figure 2a presents electrical resistance as a function of temperature for our CaVO$_3$ and SrVO$_3$ thin films grown by pulsed laser deposition (PLD) on SrTiO$_3$ (100) substrates. The film thicknesses and surface ordering were routinely monitored with single-layer accuracy by reflection high-energy electron diffraction (RHEED), see Fig. 1 in S.I.. For the thinner films (15 u.c., 10 u.c., 6 u.c., 4 u.c. CaVO$_3$) a thin 2 nm SrTiO$_3$ overlayer was deposited to protect against environmental (in air) degradation upon removal from PLD chamber. The thick 50 u.c. films are metallic at all temperatures, like bulk CaVO$_3$ and SrVO$_3$ [10,17]. When the CaVO$_3$ (CVO) film thickness is reduced a MIT occurs. The 15 u.c. and 10 u.c. films undergo a MIT at 160 K and 220 K respectively while the 6 u.c. and 4 u.c. films are insulating at all temperatures (the 4 u.c. film was too insulating to be measured using our setup). The overall behaviour is similar to that previously reported for CaVO$_3$ films [6]. Comparison of RHEED patterns recorded at room temperature after deposition of 15 u.c. CVO (metallic at room temperature) and 6 u.c. CVO (insulating at room temperature) films suggests structural transitions (possibly octahedral rotation) to play a role in the thickness-induced MIT (See S.I. Fig 3). We emphasize that this structural reconstruction was only observed for the very thin CaVO$_3$ film of 6 u.c. thickness that was capped by a 2 nm (5. u.c.) SrTiO$_3$ over-layer (compare Fig. 1 and Fig. 3 in S.I.). We ascribe an important role to the SrTiO$_3$ over-layer in inducing and/or stabilizing this structural reconstruction.

X-ray absorption (XAS) at the V L$_{2,3}$ edges presented in Figure 2b reveals an evolution from broad d-bands and large valence fluctuations in metallic thick films of CaVO$_3$ to a more localized 3d character in the thinner insulating films. XAS of the thick metallic 50 u.c. film (red line) is similar to that of a CaVO$_3$ single crystal (brown), as well as a 50 u.c. SrVO$_3$ film (dashed black). These spectra are not well reproduced by multiplet calculations because of the itinerant character of the 3d electrons [18]. With reduced film thickness the XAS spectra become sharper, the $L_3$ peak energy is reduced by ~ 0.5 eV, a shoulder appears on the $L_3$ peak and a new peak is found in the dip between $L_3$ and $L_2$ peaks. These new peaks are commonly observed in insulating $VO_2$ ($V^{4+}$, dashed grey) and $V_2O_3$ ($V^{3+}$) compounds (dashed magenta) [19]. In particular, XAS of the thinner most insulating $CaVO_3$ films (6 u.c. and 4 u.c.) is similar to that reported for $Y_{1-x}Ca_xVO_3$ ($x \geq 0.6$) that contains both $V^{4+}$ and $V^{3+}$ character [18]. The increase in $V^{3+}$ results from charge redistribution in the thinner films and is consistent with a reduced Madelung potential and hence reduced V-O covalency across the MIT. The sharpening of the $L_{2,3}$ peaks across the MIT reflects a reduction in the bandwidth of the upper Hubbard band. XAS of the 10 u.c. film in the high temperature metallic state is presented as a dashed blue curve in Figure 2b. A broadening of the spectral features is observed with respect to the low temperature insulating state but the overall response is still different from that of the metallic thick 50 u.c. $CaVO_3$ film. We undertook a series of resonant inelastic x-ray scattering (RIXS) measurements to investigate further the change in the electronic excitations across the thickness-induced and thermal MIT.

**Control of electronic bandwidth in $CaVO_3$.**

Figure 3 presents RIXS intensity maps of energy transfer versus incident energy at T = 20 K across the V $L_3$ edge. First, we discuss the energy map for the 50 u.c. $SrVO_3$ and $CaVO_3$ films presented in the panels 3a and 3b. Three fluorescence features [20], where the energy transfer disperses linearly with incident energy, are evident at higher energy transfers (marked as F1, F2, F3 in Figure 3c). These spectral signatures have been observed previously for single crystal $CaVO_3$ and were assigned to, for decreasing energy transfer, a hybridized V-O band (F3), an inter-band transition involving the lower Hubbard band (F2), and an intra-band transition involving a quasiparticle-like band (F1) [21]. Furthermore, we observe several new Raman-like modes where the energy transfer is independent of incident energy. We assign the intense Raman mode (R1) located at ~ 0.3 eV (0.5 eV for SrVO$_3$) to electron-hole pair excitations [22]. The Raman-like scattering at higher energy transfers (R2) is assigned to crystal-field excitations. The assignment of fluorescence and Raman peaks is illustrated in a schematic representation of occupied and unoccupied density of states in Figure 3e.

The RIXS response undergoes a large change as we traverse the thickness-induced MIT in CaVO$_3$. RIXS maps for insulating 10 u.c. and 4 u.c. films are presented in Figure 3c,d. Our first observation is that the spectral weight in the fluorescence features has decreased relative to the spectral weight in the Raman modes as the film thickness is decreased. This transfer of spectral weight from band to localised excitations is expected as we move from metallic to insulating behaviour. We also observe that the measured peaks become sharper and there is a reduction of the bandwidth of these electronic excitations. The reduced bandwidth of electronic excitations measured across the MIT in our RIXS experiment on CaVO$_3$ is similar to the reduced bandwidth measured using angle-resolved photoemission (ARPES) across the thickness-induced MIT in thin films of another vanadate, SrVO$_3$, that was ascribed to a dimensional crossover below 3-4 u.c. [5]. In the case of CaVO$_3$ reported here the crossover occurs at a larger thickness of 10 u.c. suggesting that both strain and dimensionality influence the MIT. Indeed it has been shown that strain can substantially affect the MIT in other systems, such as NdNiO$_3$ [23].

We quantify the reduction in the bandwidth of the electronic excitations by fitting the data and present the corresponding line cuts and fit results in Figure 4. The thickness-dependence of the low energy Raman-like RIXS peak that we assign to electron-hole pair excitations [20,22,24] is presented in Figure 4a. The elastic line has been subtracted and the corresponding fits are presented in supplementary information (S.I.). We extract from our fits the bandwidth and energy scale of the electron-hole pair excitations and present these results in Figure 4c (bottom). The bandwidth is gradually reduced from 0.73 eV in the SrVO$_3$ 50 u.c. film to 0.63 eV in the CaVO$_3$ 50 u.c. film to 0.38 eV in the insulating 6 u.c. and 4u.c. CaVO$_3$ films. The peak energy exhibits a large jump from 0.5 eV to 0.3 eV from SrVO$_3$ 50 u.c. to CaVO$_3$ 50 u.c. and then is gradually reduced to nearly 0.2 eV in the insulating 6 u.c. and 4 u.c. films.

In Figure 4b, 4c (top) we present line cuts and fit results for the fluorescence features F1 and F2 that are band excitations. We illustrate the origin of the F1 and F2 features in Figure 3 (bottom) and assign F1 to intra-band excitations involving a quasiparticle-like band and F2 to inter-band excitations involving the lower Hubbard band [21]. The separation between these two RIXS peaks is ~ 1.8 eV, the same energy separation as the low-energy optical conductivity peaks that were also interpreted as originating from intra- and inter-band excitations. Further support for this interpretation is provided by photoemission spectra that reveal a lower Hubbard band at 1.8 eV below the Fermi level [25]. The total bandwidth of our two measured RIXS peaks, determined from the first derivative of the spectra (shown in S.I.), scales with the bandwidth of the V 3d valence band [19]. Figure 4c (top) presents the results of this analysis. The total bandwidth displays a sharp drop between the metallic 50 u.c. CaVO$_3$ film (= 4 eV) and the 15 u.c. CaVO$_3$ film (= 3.5 eV) that is on the verge of being insulating (see Figure 2a). The bandwidth continues to decrease gradually to 3.2 eV in the most insulating 4 u.c. CaVO$_3$ film. A guide for the eye in Figure 4b shows the reduction in the separation between intra- and inter-band peaks, consistent with the bandwidth reduction extracted from our derivative analysis. For the 4 u.c. film the relative spectral weight in the intra-band peak substantially decreases with respect to the inter-band peak (compared to the 50 u.c. CaVO$_3$ film) as the spectral weight is transferred away from the valence band maximum. However, there is still weight in the intra-band peak even in the insulating phase indicating metallic puddles are still present even in the insulating phase [26].

**Local moment formation in CaVO$_3$ films.**

There are several Raman-like peaks marked as R2 in Figure 3 and we present line cuts through this region for different film thicknesses in Figure 5 (left). In the thick metallic 50 u.c. SrVO$_3$ film, that is nominally a perfect cubic structure with no octahedral tilting and formal valence band configuration of V 3d$^1$, we observe a broad peak centered around 1.9 eV, a similar energy scale to the t$_{2g}$-e$_g$ excitations found near 1.7 eV in the insulator YVO$_3$ with a valence electron configuration of V 3d$^2$ [27]. For the 50 u.c. CaVO$_3$ film, that has a small orthorhombic distortion and octahedral tilting, we find a large broadening and nearly flat distribution of spectral weight. A three peaks structure is measured in the 15 u.c. CaVO$_3$ film and these peaks continue to sharpen in the insulating 10 u.c., 6 u.c. and 4 u.c. CaVO$_3$ films (Figure 5 (left)). There are two possible explanations for this reconstruction: strain-induced changes in bond lengths or the formation of local magnetic moments [27]. The effect of epitaxial strain is larger in the thinner films but strain effects are expected to lift the degeneracy of the t$_{2g}$ and e$_g$ states by only a small amount (~ 100 meV) [15] compared to the large effect reported here. The additional peaks are therefore a clear signature of local moment formation from V$^{3+}$ ions allowing spin flips from a t$_{2g}^2$ high spin S=1 ground state configuration to configurations involving S=0 and the e$_g$ level [27]. We already have evidence that V$^{3+}$ ions are present in thinner CaVO$_3$ films from our XAS as described earlier and thus we can attribute the additional peak near 1 eV to a local spin flipping t$_{2g}^2$ S=0 transition at twice the Hund's coupling energy 2J$_H$ using the notation of ref. [27]. The additional peak near 2.6 eV is also captured by several overlapping local excitations on the V$^{3+}$ ion, namely by transitions to t$_{2g}^2$ with S=0 of higher crystal field symmetry and to t$_{2g}^1$e$_g^1$ with S=1 or S=0 [27]. The formation of a local moment in insulating ultra-thin CaVO$_3$ films starting from a bulk-like non-magnetic thin film is a dramatic effect, resulting from a control of electronic bandwidth and charge redistribution.

**Thermal MIT in 10 u.c. CaVO$_3$ film and the role of V-O hybridization.**

Now we draw the attention to the MIT which occurs below 220 K in the 10 u.c. CaVO$_3$ film (see Figure 2a). We present in Figure 6 the RIXS intensity as a function of energy transfer for the 10 u.c. film in the room temperature metallic phase (at 300 K) and at 20 K, deep in the insulating phase. The change in the RIXS response across the thermal MIT is qualitatively similar to the change for the thickness-induced MIT (see also Fig. 4c): the bandwidth of the electron-hole pair excitation decreases by ca. 15% along with a small softening of the peak position (Figure 6a), the crystal field levels sharpen (Figure 6b), and the total bandwidth near the valence band maximum decreases by about 5% (Figure 6c). In addition, we can quantitatively comment on the relative strength of the high-energy V 3d - O 2p hybridization band shown in Figure 6d: the intensity of this peak is reduced by 10% in the insulating phase. The reduced V-O hybridization across the thermal MIT is comparable to the relative suppression of weight in this hybridized band upon entering the thickness-induced insulating state as shown in Figure 3 (also see line cuts in S.I. Fig. 7). Thus, the same trend of reduced electronic bandwidth and reduced V-O hybridization is measured in both thermal and thickness-induced MIT, with the effects on the thickness-induced MIT being a factor 2-3 larger.

**Discussion**

The large reduction of bandwidth in our CaVO$_3$ films is comparable to that realized in the 3d$^1$ perovskite series SrVO$_3$-CaVO$_3$-LaTiO$_3$-YTiO$_3$ as shown in Figure 1. The MIT in the latter case is attributed to an increased orthorhombic distortion, and local moment formation, as the cation size is changed [16]. Obviously control of this MIT is chemically difficult as it involves substituting two elements, illustrating the usefulness of the additional parameter space enabled by thin film synthesis methods.

We can understand our results by comparing the experimental spectra for the thermal- and thickness-induced MIT. In the thermal MIT case at constant film thickness we still need to consider the epitaxial strain. Strain will induce changes in the V-O bond lengths and V-O overlap/hybridization as experimentally observed in other $3d^1$ perovskites [28]. We thus propose that a structural and/or magnetic transition is driving the thermal MIT in our 10 u.c. CaVO$_3$ film as observed in e.g. nickelate films such as NdNiO$_3$ [23]. This proposal is supported by the data presented in Figure 6b that shows a sharpening of crystal field excitations across the thermal MIT. For the thinner CaVO$_3$ films, the electron-hole bandwidth (see Figure 1 and 4 c,d) is sharply reduced moving from 10 u.c. to 6 u.c. and 4 u.c. films. Additionally XAS reveals increasing V $3d^2$ contributions (Figure 2b) and the crystal field excitations become much shaper for the thickness-induced MIT (Figure 5).

The increased contribution from V$^{3+}$ ions in the thinner films stems from charge redistribution and is connected to a reduced V-O covalence across the MIT. The reduction of the V 3d bandwidth is driven by a structural reconstruction accompanied by octahedral rotation, that is thereby also changing the V-O hybridization, manifested in an abrupt change of the RHEED patterns of the ultra-thin capped films. In this sense the "electronic localization and bandwidth reduction" as well as the "increase of V$^{3+}$ contributions" are a consequence of the thickness confinement. Oxygen vacancies generated during the growth of CVO films might partially contribute to V$^{3+}$ states, but their amount, in principle, should scale with the film thickness. However, 50 u.c. CVO film does not show more $V^{3+}$ signature in the spectra than thinner CVO films suggesting no significant influence of oxygen vacancies on the properties of the studied CVO films.

The combined effects of charge redistribution, confinement, as well as strain are important in the ultrathin $CaVO_3$ films [29]. Based on the results presented here it is difficult to distinguish which of these effects is the main player driving the thickness-induced MIT. However, we note that the MIT appears to be significantly more complex than the dimensional crossover-driven MIT reported in ultrathin $SrVO_3$ films using angle-resolved photoemission spectroscopy [5].

Our experiments reveal a bandwidth-driven MIT in $CaVO_3$ films as a function of both thickness and temperature where strain, dimensionality and charge redistribution contribute. A reduced V-O hybridization in the insulating state reveals that the MIT is of Mott-Hubbard type where a d-d gap is opened by a strong Coulomb interaction U between electrons [30]. The sharp thermal MIT observed in our 10 u.c. $CaVO_3$ film should stimulate future work on this model system. Our results are promising for the future of the field of correlated electronics showing that while increased disorder can be present in ultra-thin films, it does not inhibit the Mott physics and related spin or charge ordering. However, we want to emphasise the radical changes in electronic structure that can occur in films of reduced dimensionality, and the importance of including these effects in the rational design of new quantum materials.

**Methods**

**Experimental details.**

X-ray absorption (XAS) and resonant inelastic x-ray scattering (RIXS) experiments were carried out at the ADRESS beamline of the Swiss Light Source at the Paul Scherrer Institut [31]. All measurements were performed at grazing incidence with the x-rays incident at 15º with respect to the sample surface and with σ or π polarisation. The scattering angle was fixed at 130°. We set the spectrometer [32] in the high throughput configuration using the 1500 lines per mm variable line spacing (VLS) spherical grating [33] as well as the newly installed CCD camera that provides sub-pixel spatial resolution [34]. The beamline exit slit was 20 μm. The setup yielded a total energyresolution (full width at half maximum FWHM) of around 60 meV.

Thin films of $CaVO_3$ and $SrVO_3$ were prepared on $SrTiO_3$ (100) and $NdGaO_3$ (110) substrates respectively by pulsed laser deposition (PLD). More information can be found in the supplementary information.

**Data Availability**

The datasets generated during and/or analysed during the current study are available from the corresponding authors on reasonable request.

**Acknowledgements:**

D.M.N. and T.S. thanks A. Georges (University of Geneva) for a helpful discussion. D.M.N. acknowledges helpful discussion and analytical tools from Eugenio Paris (PSI). The work at PSI is supported by the Swiss National Science Foundation through the NCCR MARVEL and the Sinergia network Mott Physics Beyond the Heisenberg Model (MPBH). Xingye Lu acknowledges financial support from the European Community's Seventh Framework Programme (FP7/20072013) under Grant agreement No. 290605 (Cofund; PSI-Fellow). J.P. and T.S. acknowledge financial support through the Dysenos AG by Kabelwerke Brugg AG Holding, Fachhochschule Nordwestschweiz, and the Paul Scherrer Institut. J.P. acknowledges financial support by the Swiss National Science Foundation Early Postdoc Mobility fellowship Project No. P2FRP2-171824. M.D. was partially funded by the Swiss National Science Foundation within the D-A-CH programme (SNSF Research Grant 200021L 141325).

**Author Contributions:**

M.R. and T.S. conceived the project. D.M.N., M.N., T.S., M.M., C.W.S., E.V.P. and M.R. grew and characterised the thin film samples. D.M.N., X.L., J.P., M.D., V.N.S., M.R. and T.S. carried out the XAS and RIXS experiments. D.M.N., M.R., and T.S. analysed the data. S.B. and C.E. carried out the DFT calculation. D.M.N., M.R., and T.S. wrote the paper with contributions from all authors.

**Additional Information:**

The authors declare no competing financial interests. Correspondence and requests should be addressed to T.S. (thorsten.schmitt@psi.ch) and M.R. (milan.radovic@psi.ch).

**Figure Legends:**

**Figure 1 | Comparison of bandwidth change with cation substitution in bulk 3d$^1$ crystals with that realized in thin films of CaVO$_3$ by reducing film thickness.** (a and b) The crystal structure and electronic bandwidth of different 3d$^1$ perovskites RXO$_3$ is presented. For R = Sr and X =V a cubic structure is reported. When R is substituted to R = Ca an orthorhombic distortion occurs changing the electronic bandwidth and local V site symmetry. This results in a more correlated metallic state. When R is substituted to R = La and X substituted to X = Ti (to maintain 3d$^1$ configuration), the distortion is further increased and a metal-insulator transition occurs (MIT). The MIT is accompanied by long range antiferromagnetic order that can be switched to ferromagnetic by substituting R to R = Y. (c and d) A comparable reduction in bandwidth is found when the MIT in CaVO$_3$ is controlled by film thickness as described in this manuscript.

**Figure 2 | Reduced electronic bandwidth and increased V$^{3+}$ character across the metal-insulator transition in thin film CaVO$_3$. (a)** Resistance as a function of temperature for the samples indicated. Arrows indicate where the slope of the resistance curve reverses sign for the 15 u.c. and 10 u.c. films. **(b)** X-ray absorption (XAS) as total fluorescence yield for a CaVO$_3$ crystal and films with thicknesses as indicated (solid lines) for π-polarised light at a temperature of 20 K. XAS of the 10 u.c. CaVO$_3$ film at room temperature is shown as a dashed blue curve. Also shown is the XAS for a 50 u.c. SrVO$_3$ film (dashed black) and VO$_2$ crystal (dashed grey) and V$_2$O$_3$ crystal (dashed magenta).

**Figure 3 | Overview of resonant inelastic x-ray scattering (RIXS) measurements across a thickness-induced MIT in vanadate thin films. (top)** X-ray absorption total fluorescence yield for samples indicated **(a,b,c,d).** RIXS intensity maps of energy transfer versus incident energy for

different film thickness across the thickness-induced MIT. CaVO$_3$ and SrVO$_3$ metal films are 50 unit cells thick. **(e)** Illustrative sketch of the origin of the Raman and fluorescence peaks measured using RIXS. The sketch is drawn on the total density of states (DOS) and partial DOS for V 3d e$_g$ (red) and V 3d t$_{2g}$ (blue) calculated using density functional theory.

**Figure 4 | 40 % bandwidth reduction across the thickness-induced MIT in CaVO$_3$ thin films. (a,b)** Resonant inelastic x-ray scattering (RIXS) intensity as a function of energy transfer for the films indicated for an incident energy of 520 eV (in the dip between L$_3$ and L$_2$ peaks). SrVO$_3$ and CaVO$_3$ are 50 u.c. films and also shown are 15 u.c., 10 u.c., 6 u.c. and 4 u.c. CaVO$_3$ films. All data were recorded at a temperature of 20 K, and normalized to the total integrated intensity and offset by additive constants for clarity except data in (a) that were multiplied by small factors away from unity for clarity. **(c) (top)** V 3d bandwidth of the two features F1 and F2 for different films indicated, as extracted from the data presented in (b); fits for extracting the valence band bandwidth are shown in the supplementary information (S.I.). **(bottom)** Bandwidth and peak position of the electron-hole pair excitations for different films indicated, extracted from the fits shown in S.I. The empty red square, diamond, circle represent bandwidth and peak positions extracted from measurements of the 10 u.c. film in the metallic phase at 300 K presented in Fig. 6.

**Figure 5 | Reconstruction of crystal field levels across the thickness-induced MIT. (left)** Resonant inelastic x-ray scattering (RIXS) intensity as a function of energy transfer for the films indicated for an incident energy of 520 eV (in the dip between L$_3$ and L$_2$ peaks). SrVO$_3$ and CaVO$_3$ are 50 u.c. films and also shown are 15 u.c., 10 u.c., 6 u.c. and 4 u.c. CaVO$_3$ films. All data were recorded at a temperature of 20 K, and normalized to the total integrated intensity and offset by additive constants for clarity. Labels indicate the different allowed crystal field excitations for V$^{4+}$ and V$^{3+}$

ions. **(right)** Schematic of the change in the electronic levels of the V 3d electrons due to reconstruction of the local V site environment and valence occupation.

**Figure 6 | Resonant inelastic x-ray scattering across the thermal metal-insulator transition in a 10 u.c. CaVO$_3$ film. (a,b,c,d)** Resonant inelastic x-ray scattering (RIXS) intensity as a function of energy transfer at incident photon energy of 520 eV for a 10 u.c. CaVO$_3$ at 20 K and 300 K. **(inset of a)** Same with elastic line subtracted.

**Figures**

**FIGURE 1**

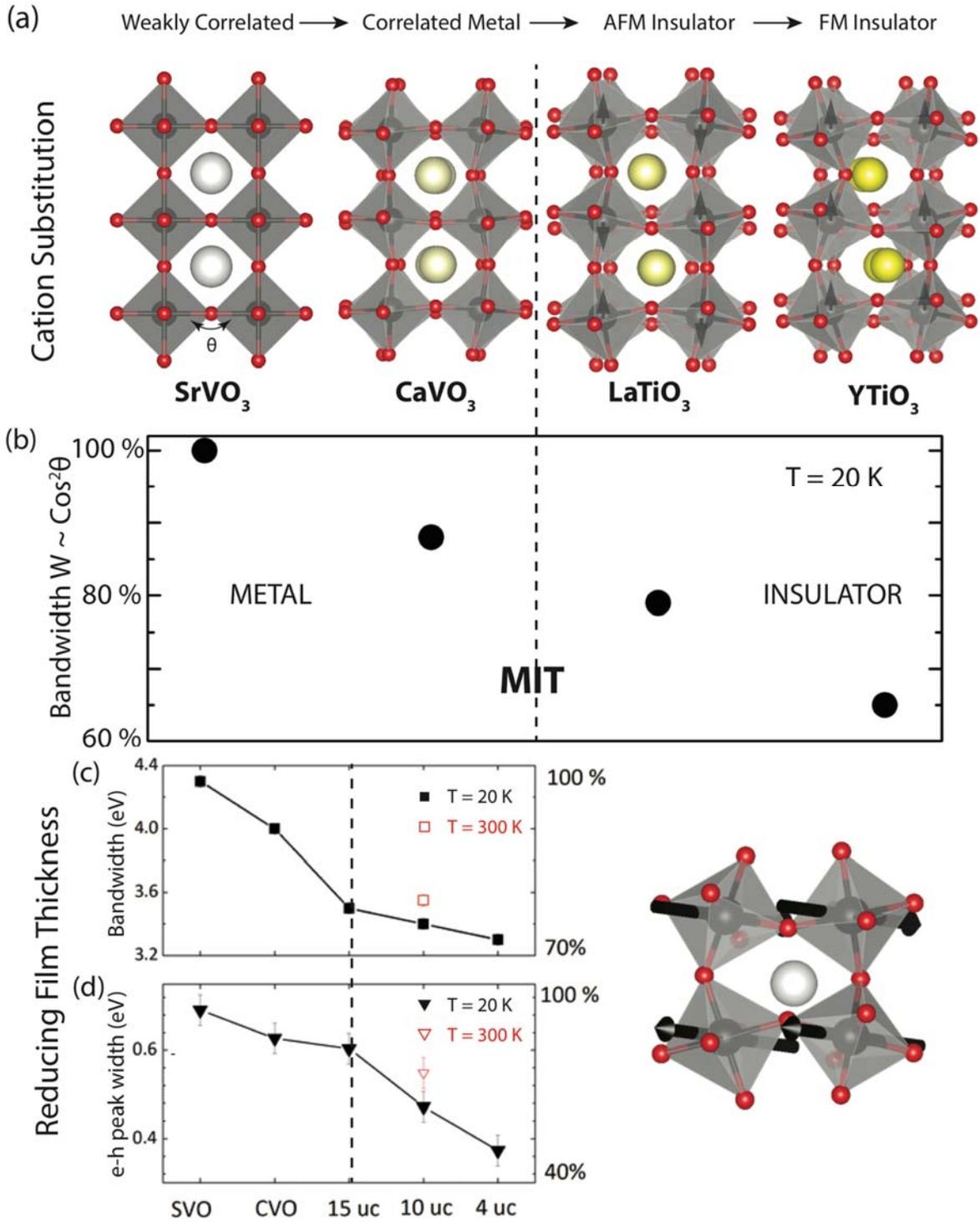

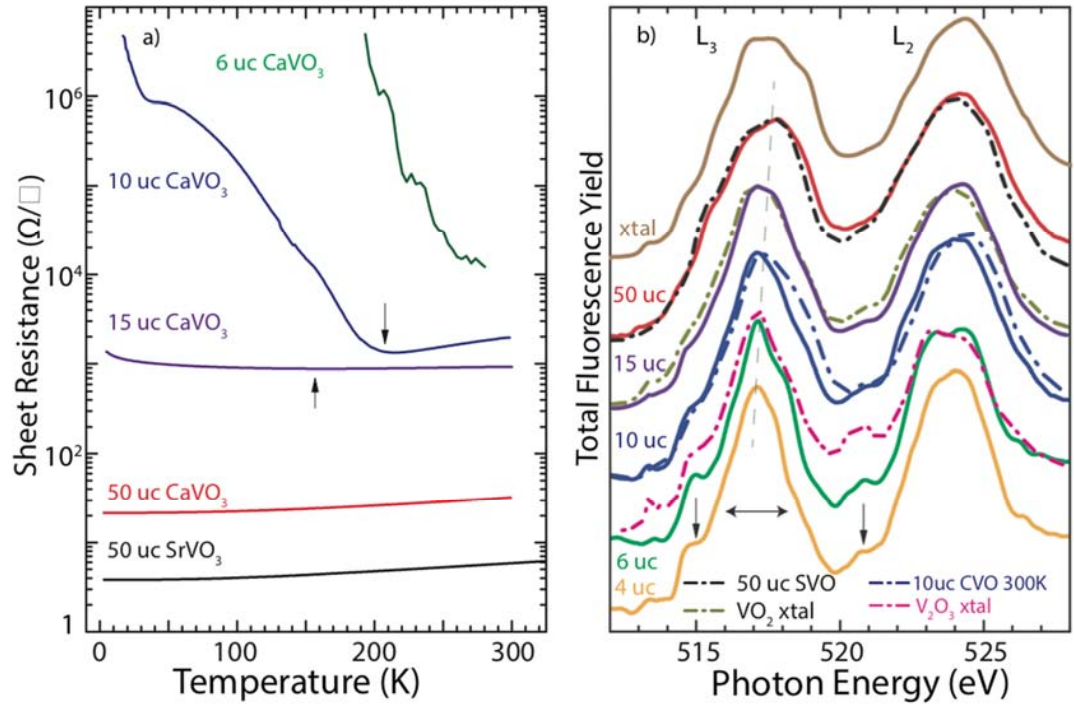

**FIGURE 3**

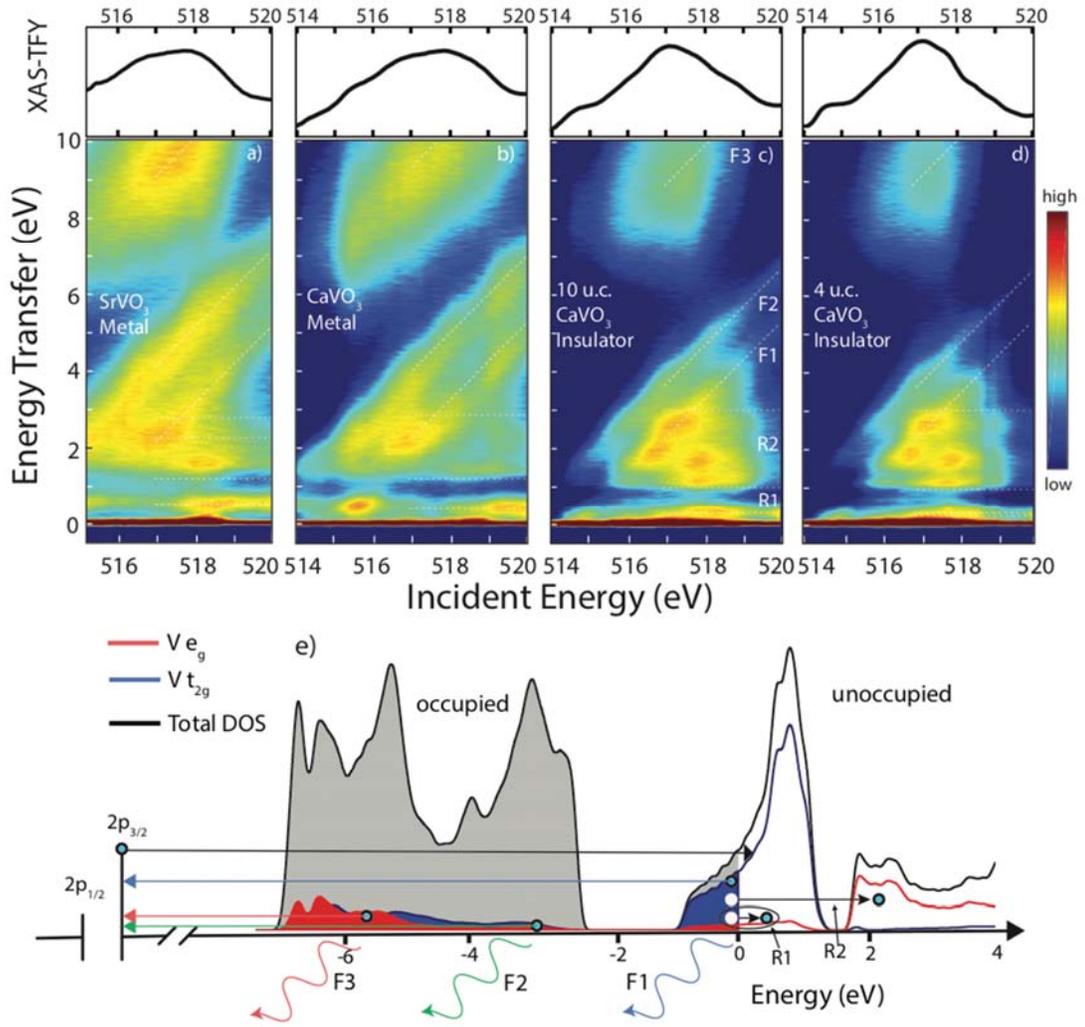

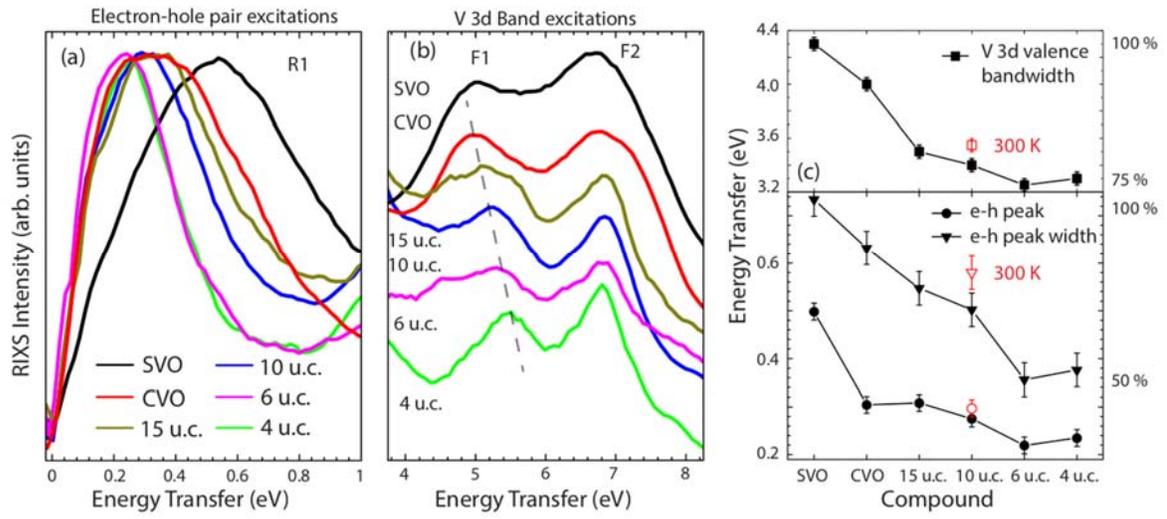

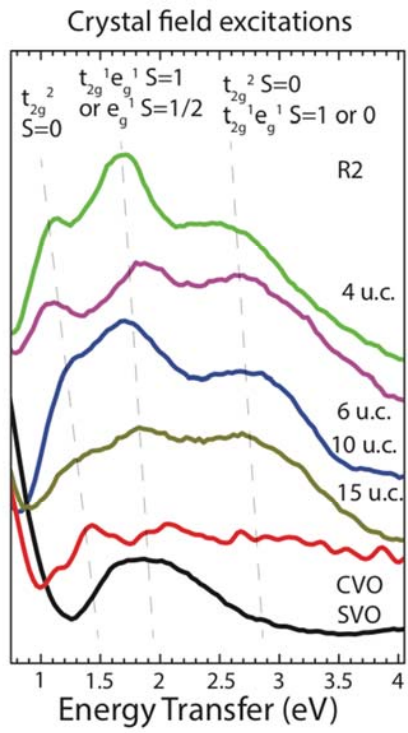
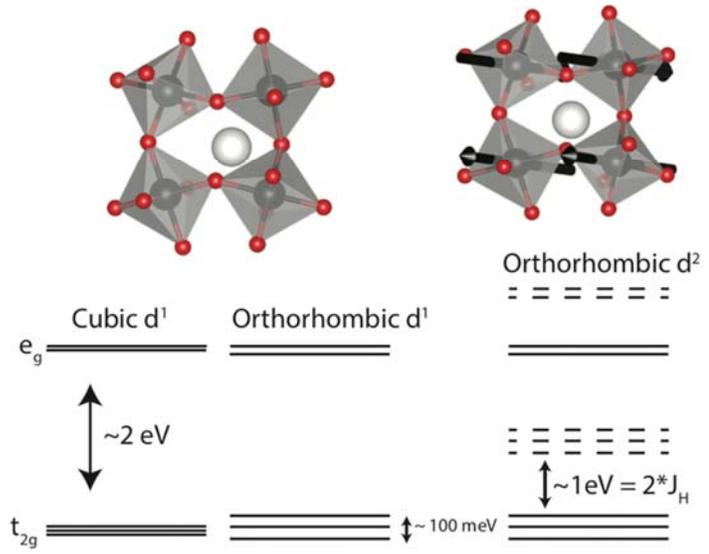

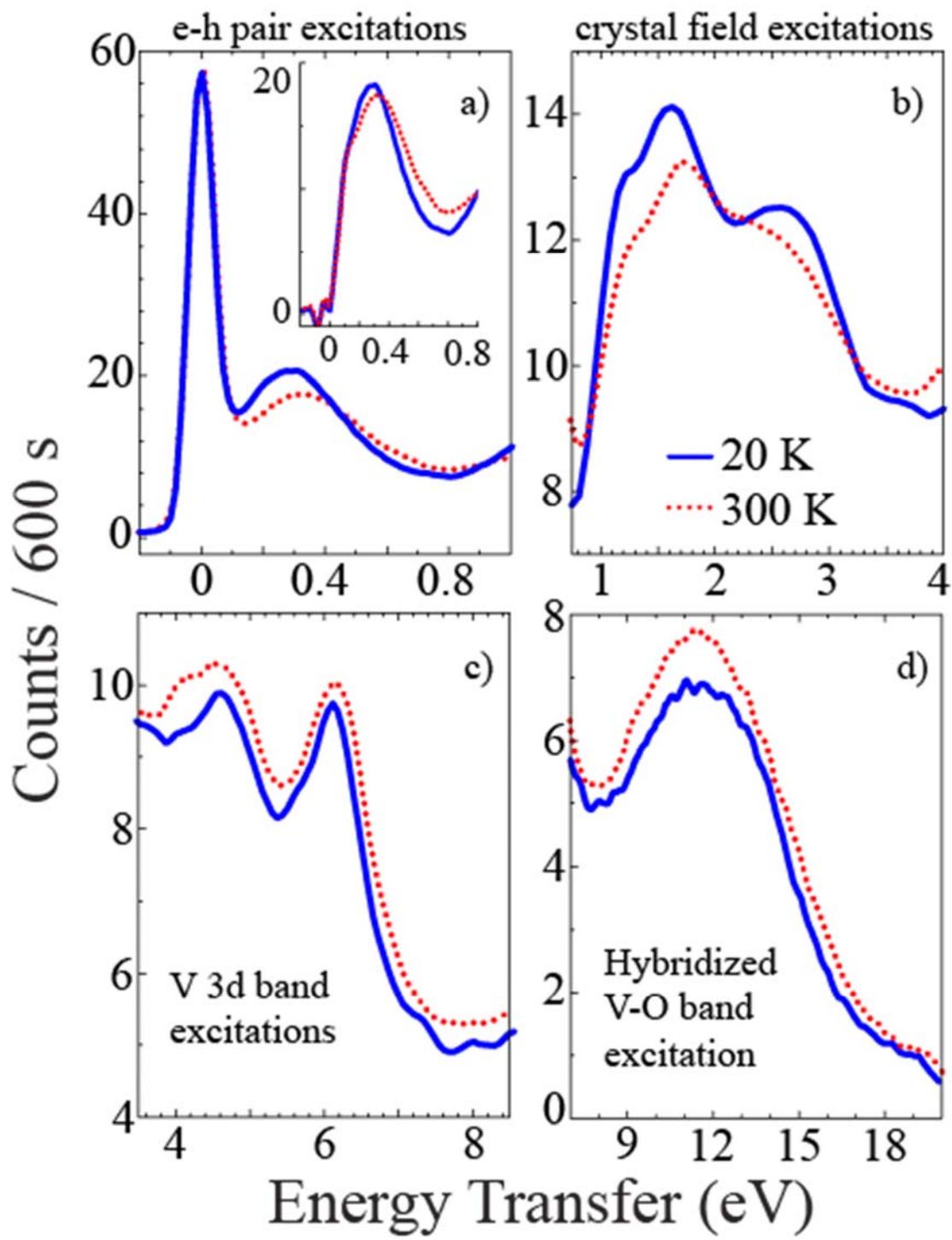